\documentclass[preprint,superscriptaddress,a4paper,showpacs,showkeys,11pt,nofootinbib]{revtex4}
\usepackage{graphicx} 
\usepackage{graphicx}
\usepackage{amsmath}
\usepackage{booktabs}
\usepackage{siunitx}
\usepackage{url}
\usepackage{hyperref}
\newcommand{\rr}{\mbox{\boldmath $r$}}

\def\pom{{I\!\!P}}

\newcommand{\dd}{\, \mathrm{d}}
\newcommand{\rb}{\mbox{\boldmath $b$}}

\begin{document}
\title{Photoproduction of heavy vector mesons in peripheral $PbPb$ collisions \\ at the Large Hadron Collider}

\author{Pedro E. A.  da {\sc Costa}}
\email{pazevedo138@gmail.com}
\affiliation{Departamento de F\'isica, Universidade do Estado de Santa Catarina, 89219-710 Joinville, SC, Brazil.}

\author{Andr\'e V. {\sc Giannini}}
\email{AndreGiannini@ufgd.edu.br}
\affiliation{
Federal University of Grande Dourados, 
Faculty of Exact Sciences and Technology,
Zip code 364, 79804-970, Dourados, MS, Brazil}
\affiliation{Departamento de F\'isica, Universidade do Estado de Santa Catarina, 89219-710 Joinville, SC, Brazil.}

\author{Victor P. {\sc Gon\c{c}alves}}
\email{barros@ufpel.edu.br}
\affiliation{Institute of Physics and Mathematics, Federal University of Pelotas (UFPel), \\
  Postal Code 354,  96010-900, Pelotas, RS, Brazil}

\author{Bruno D. {\sc Moreira}}
\email{bduartesm@gmail.com}
\affiliation{Departamento de F\'isica, Universidade do Estado de Santa Catarina, 89219-710 Joinville, SC, Brazil.}

\begin{abstract}
A comprehensive analysis of the photoproduction of $J/\Psi$ and $\Upsilon$ mesons in peripheral  $PbPb$ collisions at the center - of - mass energies of the Large Hadron Collider (LHC) is performed, considering distinct assumptions for the modeling of the nuclear photon flux, photon - nucleus cross - section, overlap function and dipole - proton scattering amplitude. The comparison of these predictions with the ALICE data is also performed. Our results indicate that a detailed analysis of the production of both mesons  will be very useful to improve the description of photon - induced processes in peripheral collisions.
\end{abstract}


\maketitle

\section{Introduction}

The study of ultraperipheral heavy - ion collisions (UPHICs) became a reality in the last decade, which has allowed us to improve our understanding of photon - induced interactions and probe distinct final states (For reviews see Ref. \cite{upc}). In particular, the exclusive vector meson photoproduction in UPHICs  were largely investigated over the last years, motivated by the possibility to constrain the description of the QCD dynamics at high energies \cite{klein,gluon,Frankfurt:2001db},  and improve our understanding of the quantum 3D imaging of the partons inside the protons and nuclei \cite{upc2}. Currently, distinct approaches are able to describe the UPHIC data, but the situation is expected to be improved  in the forthcoming years with the releasing of new data.

Over the last years, the measurements of the $J/\Psi$ yields at very low transverse momentum in peripheral collisions by the  STAR,  ALICE and LHCb Collaborations \cite{ALICE:2015mzu,STAR:2019yox,LHCb:2021hoq,ALICE:2022zso,Massacrier:2024fgx,Bize:2024ros} have demonstrated  that the  photoproduction of $J/\Psi$ mesons  becomes larger than the contributions associated with the vector meson production in hadronic interactions. As a consequence, peripheral collisions can also considered an alternative way to improve our understanding about the QCD dynamics and nuclear structure. Such a possibility has motivated the proposition of distinct approaches to treat photon - induced interactions in peripheral collisions \cite{Klusek-Gawenda:2015hja,Zha:2017jch,GayDucati:2017ksh,Shi:2017qep,Zha:2018ytv,GayDucati:2018who}. In this paper, we will review some of these models, and perform a comprehensive study of the photoproduction of $J/\Psi$ and $\Upsilon$ mesons in peripheral $PbPb$ collisions at the LHC energies.  In particular, we will employ distinct models for the nuclear photon flux, photon - nucleus cross - section, overlap function and dipole - proton scattering amplitude. Our goal is to estimate the current theoretical uncertainty in the treatment of peripheral collisions, to compare the predictions with the existing data and provide predictions that could be compared with the future measurements.

This paper is organized as follows. In the next Section, we present a brief review of the formalism needed to describe the  photoproduction of heavy vector meson in peripheral $PbPb$ collisions at the LHC. We will review the distinct models used in the literature for the treatment of  nuclear photon flux and  photon - nucleus cross section in peripheral collisions.  In particular, the different ingredients needed to describe the nuclear vector meson photoproduction in the color dipole formalism will be discussed. In Section \ref{sec:results} we will present our predictions for the rapidity distributions associated with the photoproduction of $J/\Psi$ and $\Upsilon$ mesons, derived considering different models for the effective nuclear photon flux, dipole - proton scattering amplitude, overlap function and  photon - nucleus cross - section. Moreover, a comparison with the ALICE data will be performed. Finally, in Section \ref{sec:summary} we will summarize our main conclusions. { For completeness of our study, three appendices are included, where we discuss the relation between the centrality of the collision and the corresponding  impact parameter, and  present our predictions for ultraperipheral collisions, as well as for peripheral collisions characterized by a centrality smaller than 50\%.}

\begin{figure}[t]
\includegraphics[scale=0.6]{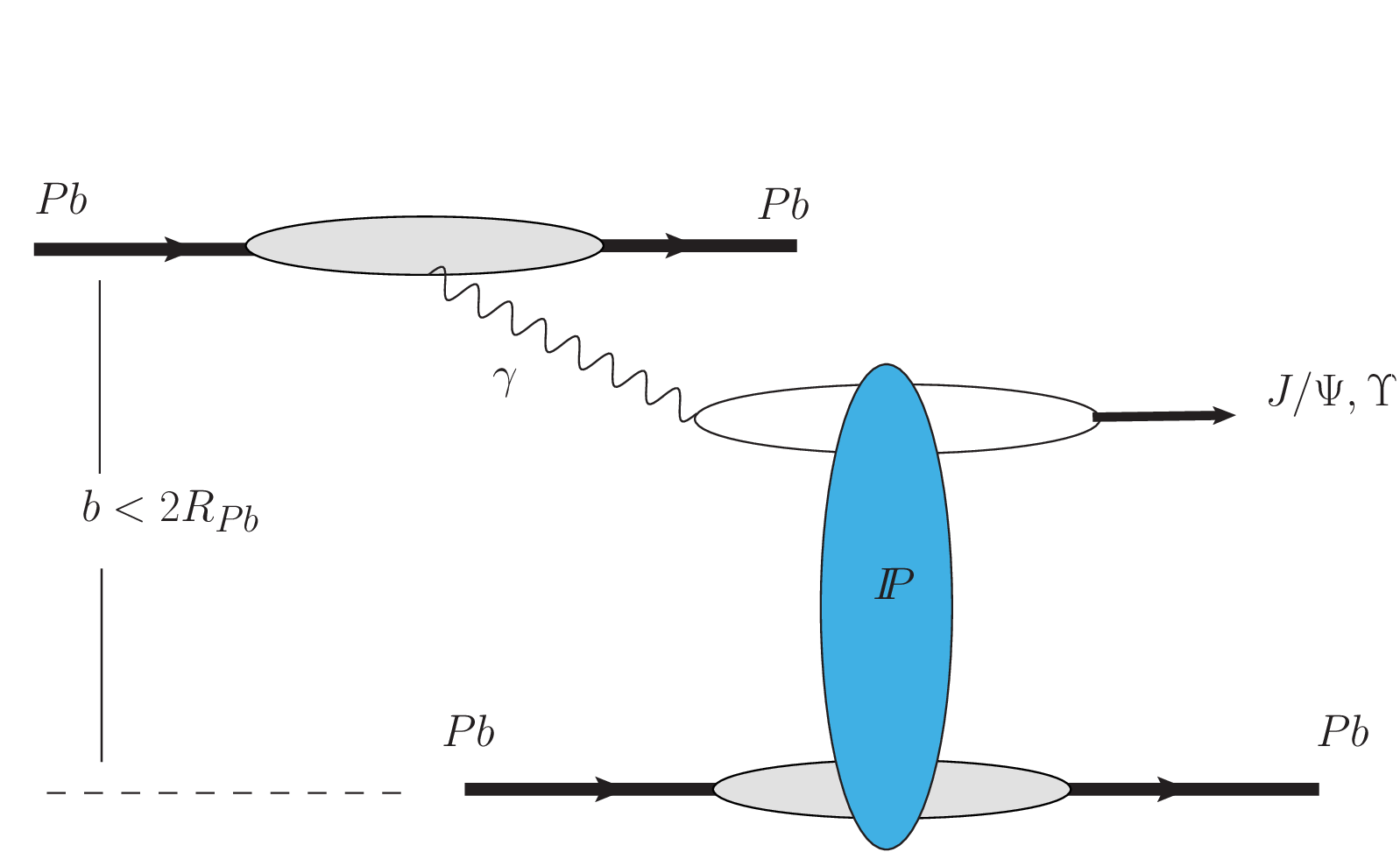}
\caption{Photoproduction of heavy vector mesons in peripheral $PbPb$ collisions.}
\label{Fig:diagram}
\end{figure}

\section{Formalism}
The photoproduction of heavy vector mesons in peripheral $PbPb$ collisions is represented in Fig. \ref{Fig:diagram}.  
In the equivalent photon approximation (EPA) \cite{epa}, the associated cross - section, characterized by an impact parameter $\rb$, can be expressed in terms of  the equivalent photon spectrum associated with one of the ions and by the  production cross section that described the interaction of the photon with the other ion. Considering that both ions can act as photons sources and nuclear targets, one has that differential cross - section for the vector meson photoproduction in an $AB$ collision will be given by
\begin{eqnarray}
\frac{d\sigma \,\left[A B \rightarrow   A \otimes  V \otimes B\right]}{d^2\rb \, dy_V} = \omega_A N_{A}(\omega_A,b)\,\sigma_{\gamma B \rightarrow V \otimes B}\left(\omega_A \right) + \omega_B N_{B}(\omega_B,b)\,\sigma_{\gamma A \rightarrow V \otimes A}\left(\omega_B \right) \,\,,
\label{dsigdy}
\end{eqnarray}
where $b = |\rb|$, $y_V$ is the rapidity  of the vector meson in the final state, $\omega_A$ ($\omega_B$) is energy of the photon emitted by the nucleus $A$ ($B$). One has that $\omega_A = (m_V/2)e^{y_V}$ and $\omega_B = (m_V/2)e^{-y_V}$, with $m_V$ the mass of the vector meson.
Moreover, $\sigma_{\gamma A_i \rightarrow V \otimes A_i}$ is the cross - section for the production of a vector meson in a photon - nucleus interaction, and the symbol
$\otimes$ represents that the photon - nucleus interaction was mediated by a color singlet object, usually denoted Pomeron $\pom$, and  that a rapidity gap is expected to be present in the final state. { The photon spectrum, $N(\omega,b)$, can be expressed in terms  in terms of the charge form factor $F(q)$ as follows \cite{upc}
\begin{eqnarray}
 N(\omega,b) = \frac{Z^{2}\alpha}{\pi^2}\frac{1}{b^{2} v^{2}\omega}
\cdot \left[
\int u^{2} J_{1}(u) 
F\left(
 \sqrt{\frac{\left( \frac{b\omega}{\gamma_L}\right)^{2} + u^{2}}{b^{2}}}
 \right )
\frac{1}{\left(\frac{b\omega}{\gamma_L}\right)^{2} + u^{2}} \mbox{d}u
\right]^{2} \,\,,
\label{Eq:fluxo0}
\end{eqnarray}
where $\alpha$ is the electromagnetic coupling constant, $\gamma_L$ is the Lorentz factor and $v$ is the nucleus velocity. In what follows, we will estimate the photon flux using the realistic
form factor, which corresponds to the Wood - Saxon distribution and is the Fourier transform of the charge density of the nucleus, being analytically expressed by
    \begin{equation}\label{realistico}
	F(q^2) = \dfrac{4\pi\rho_0}{Aq^3}\big[\sin(qR_A) - qR_A\cos(qR_A)\big]\left[\dfrac{1}{1 + q^2 a^2}\right].
    \end{equation}
with \( R_A = 6.62 \) fm, \( a = 0.549 \) fm and  \( \rho_0 = 0.1603 \,\,\text{fm}^{-3} \) for a lead nucleus~\cite{DeJager:1974liz,Bertulani:2001zk}.

     \begin{figure}[t]
 	\centering
 	\includegraphics[width=0.45\linewidth]{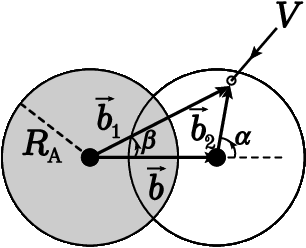}
        \caption{Transverse - plane view of the vector meson photoproduction in a peripheral $AB$ collisions.}
        \label{colisao_geometria}
    \end{figure}

In our analysis, we will focus on peripheral collisions, where $b < R_{A} + R_{B}$ and one has the overlap between the incident ions. A transverse - plane view of the heavy vector meson photoproduction in a peripheral $AB$  collision  will be characterized by three vectors in the transverse - plane (See Fig. \ref{colisao_geometria}): $|\rb|$ is the impact parameter between the nuclei $A$ and $B$, $|\rb_1|$ is the distance from the center of the emitting nucleus $A$ to the point at which the emitted photon interacts with the nucleus $B$ and produces the vector meson $V$, and $|\rb_2| = |\rb_1 - \rb|$ is the distance from the center of the nucleus $B$ to that same point. One has that ${b}_2^2 = {b}_1^2 + {b}^2 - 2b_1 b \cos(\beta)$.  
As the  treatment of the  vector meson photoproduction in peripheral collisions is still an open question, in this paper we will  consider the distinct  geometrical approaches proposed in Refs. \cite{Klusek-Gawenda:2015hja, GayDucati:2017ksh}   and will consider that the associated cross - section can be estimated  by modificating the nuclear photon flux and/or the photon - nucleus cross - section. In what follows, we will discuss the models that will be considered in our analysis assuming the nucleus $A$ as the photon emitter and the nucleus $B$ as the target. The contribution of the second term in Eq. (\ref{dsigdy}) can be directly derived by performing the transformations $A \longleftrightarrow B$ and $\rb_1 \longleftrightarrow
 \rb_2$.

\subsection{Effective nuclear photon fluxes}
In our analysis we will consider the models proposed in Refs. \cite{Klusek-Gawenda:2015hja, GayDucati:2018who}  to extend the treatment of the nuclear photon flux for peripheral heavy - ion collisions. The simplest approach is to assume that the nuclear photon flux is not modified in peripheral collisions and derive the predictions using  Eq. (\ref{Eq:fluxo0}), denoted $N^{(0)}_A(\omega, b)$ hereafter, as input of calculations.  Another possibility is that the photon flux is modified in peripheral collisions, and that this modification can be implemented considering geometrical aspects of the colllision and additional assumptions about the nuclear regions that contribute for the cross - section, as suggested in Refs. \cite{Klusek-Gawenda:2015hja,GayDucati:2018who}. In particular, Ref. \cite{Klusek-Gawenda:2015hja}    has proposed two distinct approaches. The first one, considers that the production of vector mesons only occurs inside the nuclear target, which is equivalent to assume that  
    \begin{equation}
N^{(1)}_A(\omega, b) = \int N^{(0)}_A(\omega, {b}_1) \frac{\theta(R_B - b_2)}{\pi R_B^2} \,d^2b_1\,\,,
\label{Eq:flux1} 
\end{equation}
where the theta function ensures that the photon emitted by the nucleus $A$ hits the nucleus $B$. It is important to emphasize that in this model, the overlapping region between the incident ions is included as a possible interaction region. As the photoproduction of vector mesons in this region can be strongly affected by the presence of a nuclear or hot medium, the authors of Ref. \cite{Klusek-Gawenda:2015hja} have also proposed an alternative model, where this overlapping region is excluded in the modeling of the effective nuclear photon flux. In this case, the effective nuclear photon flux is given by
    \begin{equation}
	N^{(2)}_A(\omega, b) = \int N_A^{(0)}(\omega, {b}_1) \frac{\theta(R_B - b_2) \times \theta(b_1 - R_A)}{\pi R_B^2} d^2b_1 \,\,.
    \label{Eq:flux2} 
    \end{equation}

    Finally, we also will  consider the model proposed in Ref. \cite{GayDucati:2018who}, where the constant factor $\pi R_B^2$ in the denominator of Eq.  (\ref{Eq:flux2}) is generalized by an effective area, which depends on the impact parameter of the collision. In this case, the effective nuclear photon flux is  given by \cite{GayDucati:2017ksh}
    \begin{equation}
	N^{(3)}_A(\omega, b) = \int N_A^{(0)}(\omega, b_1) \frac{\theta(R_B - b_2) \times \theta(b_1 - R_A)}{A_{\text{eff}}(b)} \,d^2b_1 \,\,,
    \label{Eq:flux3} 
    \end{equation}  
with   
\begin{equation}
	A_{\text{eff}}(b) = R_B^2 \left[\pi - 2\arccos\left(\frac{b}{2R_B}\right)\right] + \frac{b}{2} \sqrt{4R_B^2 - b^2} \,\,.
\end{equation}

\subsection{Photoproduction of vector mesons in the color dipole approach}

In the color dipole approach,  the cross - section for the photoproduction of vector mesons in coherent $\gamma A$ interaction can be expressed as follows~\cite{Kopeliovich:2001xj}
\begin{equation}
\sigma_{\gamma B \rightarrow V B}(W) = \int d{^2\rb_2}\,  \left\{ \int \int dz \, d^2\rr  [\psi^*_V(r,z)\psi(r,z)]_T\, \mathcal{N}_B(x, \rr, \rb_2) \right\}^2\,\,,
\end{equation} 
where $W$ is photon - nucleus center - of - mass energy, $(\Psi^{V*}\Psi)_{T}$ denotes the overlap of the transverse photon and vector meson wave functions (see below), and  $z$ $(1-z)$ is the
longitudinal momentum fractions of the quark (antiquark). Moreover, 
 ${\cal{N}}_{B} (x,\rr,\rb)$ denotes the non-forward scattering  amplitude of a dipole of size $\rr$ on the target $B$, which is  directly related to  the QCD dynamics \cite{hdqcd}.
Following Ref. \cite{run2}, we will estimate ${\cal{N}}_B$  assuming the Glauber-Gribov (GG) formalism~\cite{glauber,gribov,mueller,Armesto:2002ny}, which predicts that
\begin{eqnarray}
{\cal{N}}_B(x,\rr,\rb) =  1 - \exp \left[-\frac{1}{2}  \, \sigma_{dp}(x,\rr^2) \,T_B(\rb)\right] \,,
\label{enenuc}
\end{eqnarray}
where the nuclear profile function $T_B(\rb)$ is described by a Wood-Saxon distribution. The dipole-proton cross section, $\sigma_{dp}$, is expressed in terms of the dipole-proton scattering amplitude as follows
\begin{eqnarray}
\sigma_{dp}(x,\rr^2) = 2 \int d^2\rb_p \, {\cal{N}}_p(x,\rr,\rb_p) \,,
\end{eqnarray}
with  $\rb_p$ being the impact - parameter for the dipole-proton interaction.

In our analysis, we will consider three distinct models for the description of $\sigma_{dp}$, two that take into account of nonlinear effects on the QCD dynamics, derived considering distinct approaches  (bCGC \cite{KMW} and  IP-SAT models), and one that disregard these effects  (IPnon-SAT  model). In what follows, we present a brief review of these models. 

The impact parameter Color Glass Condensate (bCGC) model was proposed in
Refs. \cite{KMW,Watt_bCGC} and  interpolates two analytical 
solutions of well known evolution equations: the solution of the BFKL equation near  the 
saturation regime and the solution of the  Balitsky-Kovchegov 
equation deeply inside the saturation regime. This model assumes that the saturation scale  depends on the impact parameter, with the dipole - proton scattering amplitude being given by \cite{KMW,Watt_bCGC} 
\begin{eqnarray}
{\cal N}_p(x,\rr,\rb_p) = \left\{  \begin{array}{l}
{\cal N}_{0}\left( 
\frac{rQ_{s}}{2}
\right )^{2[\gamma_{s}+(1/(\kappa \lambda Y))\ln (2/rQ_{s})]} 
\,\,\, , rQ_{s} \leq 2\\
1- e^{-A \ln^{2}(BrQ_{s})} \,\,\,\,\,\,\,\,\,\,\,\,\,\,\,\,\,\,\,\,\,\,\,\,\,\,\,\,\,\,\,\,\,\,\,\,\, , rQ_{s} > 2
\end{array} \right.  
\label{iim}
\end{eqnarray}
where  $Y=\ln(1/x)$  and
\begin{eqnarray}
  Q_{s} \equiv Q_{s}(x,b_p) = 
\left(
\frac{x_{0}}{x}
 \right )^{\lambda/2} \left[
\exp \left(
-\frac{b_p^{2}}{2B_{CGC}}
 \right )
 \right ]^{1/(2 \gamma_{s})}
\label{qs_bcgc}
\end{eqnarray}
is the saturation scale of this model. Moreover, the coefficients $A$ and $B$ are determined 
by the continuity condition of ${\cal N}$ and its derivative and are given by
\begin{eqnarray}
 A &=& - \frac{{\cal N}_{0}^{\,2}\gamma_{s}^{2}}
{(1-{\cal N}_{0})^{2} \ln (1-{\cal N}_{0})} \,\,\, , \\
B &=& \frac{1}{2}(1-{\cal N}_{0})^{-(1-{\cal N}_{0})/({\cal N}_{0}\gamma_{s})}  .
\end{eqnarray} 
The free parameters were fixed by fitting the HERA data. Here we have used the updated 
parameters from Ref. \cite{Xie:2018tng}. 
Moreover, we will consider the  IP-SAT model \cite{ipsat1,ipsat3,ipsat4} that incorporates the saturation effects via the Glauber - Mueller approximation \cite{mueller}, assuming an eikonalized 
form for ${\cal N}_p$  that depends on a gluon distribution evolved via DGLAP equation.   In the IP-SAT model, the dipole - proton scattering amplitude is  given 
by 
\begin{eqnarray}
 {\cal N}_p(x,\mbox{\textbf{\textit{r}}},\mbox{\textbf{\textit{b}}}) = 
 1 - \exp \left[-
\frac{\pi^{2}r^{2}}{2 N_{c}} \alpha_{s}(\mu^{2}) \,\,xg\left(x, \mu_{0}^{2} + \frac{C}{r^{2}}  
\right)\,\, T_p(\rb) 
 \right] ,
 \label{ipsat}
\end{eqnarray}
and  a Gaussian profile
\begin{eqnarray}
T_p(\rb) = \frac{1}{2\pi B_{p}}  
\exp\left(-\frac{\rb^{2}}{2B_{p}} \right) 
\end{eqnarray}
is assumed for the proton density profile.
In Ref. \cite{ipsat_heikke}, { the authors have assumed  $B_p = 4$ GeV$^{-2}$ 
based on a fit of the HERA exclusive $J/\Psi$ production data.
The initial gluon distribution, evaluated at the scale $\mu_{0}^{2}$, is taken to be 
\begin{eqnarray}
xg(x,\mu_{0}^{2}) =  A_{g}x^{-\lambda_{g}} (1-x)^{6} .
\end{eqnarray}
The free parameters of this model are fixed by a fit of HERA data \cite{ipsat_heikke}.  Finally, we will also consider a linearized  version of the IP-SAT model, usually denoted IPnon-SAT model, which disregard these non-linear effects and has a dipole - proton scattering amplitude given by \cite{ipsat_heikke}
\begin{eqnarray}
 {\cal N}_p(x,\mbox{\textbf{\textit{r}}},\mbox{\textbf{\textit{b}}}) = 
\frac{\pi^{2}r^{2}}{2N_{c}} \alpha_{s}(\mu^{2}) \,\,xg\left(x, \mu_{0}^{2} + \frac{C}{r^{2}}  
\right)\,\, T_p(\rb) \, \,.
 \label{ipnonsat}
\end{eqnarray}
Our analysis will be performed using the parameters obtained in Ref. \cite{ipsat_heikke}.

Another important ingredient in the calculation of the nuclear photoproduction of vector mesons in the color dipole approach is the overlap function $(\Psi^{V*}\Psi)_{T}$. 
In what follows, we will consider the Boosted Gaussian (BG) and the Gaus-LC (GLC) models \cite{wflcg,nnpz,sandapen,ipsat1}. These models assume that the vector meson  is predominantly a quark-antiquark state and  that the spin and polarization structure is the same as in the  photon. It implies that  the overlap between the photon and the vector meson wave function, for the transversely polarized  
case, is given by (For details see Ref. \cite{KMW})
\begin{eqnarray}
(\Psi_{V}^* \Psi)_T = \hat{e}_f e \frac{N_c}{\pi z (1-z)}\left\{m_f^2K_0(\epsilon r)\phi_T(r,z) -[z^2+(1-z)^2]\epsilon K_1(\epsilon r) \partial_r \phi_T(r,z)\right\} \,\,,
\end{eqnarray}
where $ \hat{e}_f $ is the effective charge of the vector meson, $m_f$ is the quark mass, $N_c = 3$, $\epsilon^2 = z(1-z)Q^2 + m_f^2$ and $\phi_T(r,z)$ define the scalar part of the  vector meson wave function. The function $\phi_T(r,z)$ is model dependent.
In the BG model, the function $\phi_T(r,z)$ is given by
\begin{eqnarray}
\phi_T(r,z) = N_T z(1-z) \exp\left(-\frac{m_fR^2}{8z(1-z)} - \frac{2z(1-z)r^2}{R^2} + \frac{m_f^2R^2}{2}\right) \,\,.
\end{eqnarray}
 In contrast, in the GLC model, it is given by
\begin{eqnarray}
\phi_T(r,z) = N_T [z(1-z)]^2 \exp\left(-\frac{r^2}{2R_T^2}\right) \,\,.
\end{eqnarray}
The parameters $N_T$, $R$ and $R_T$ are  determined by the normalization condition of the wave function and by the decay width. In Table \ref{Tab:parametrosGLC}   we present the parameters associated with the GLC model, derived assuming the distinct values for heavy quark mass used by the different dipole models and the more recent values for the decay width of the $J/\Psi$ and $\Upsilon$ mesons  \cite{ParticleDataGroup:2024cfk}. On the other hand, for the BG model, we will assume the parameters derived in Refs. \cite{ipsat_heikke,Cepila:2025rkn}. 

\begin{table}[t]
	\centering
	\renewcommand{\arraystretch}{1.3}
	\begin{tabular}{ccccccc}
		\hline
		\hline
		Meson  & $\hat{e}_f$ & $M_V$ [GeV] & $f_V$  & $m_f$ [GeV] & $N_T$ & $R_T^2$ [GeV$^{-2}$] \\
		\hline
		\hline
		$J/\psi$ & $2/3$ & $3.097$ & 0.277 & $1.27$ & $1.4532$ & $5.5175$ \\
		\hline
		\hline
		$J/\psi$ & $2/3$ & $3.097$& 0.277 & $1.3528$ & $1.3137$ & $6.0204$ \\
		\hline
		\hline
		$J/\psi$ & $2/3$ & $3.097$& 0.277 & $1.3504$ & $ 1.3175$ & $6.0052$ \\
		\hline
		\hline
		\(\Upsilon\)& \(1/3\) & \(9.46\) & \(0.234\) & \(4.18\)  & \(0.7654\) & \(1.9211\)\\
			\hline
		\hline
	\end{tabular}
	\caption{Parameters of the GLC vector meson wave functions derived considering the values for the charm and bottom mass used in the distinct models for the dipole - proton scattering amplitude. }
    \label{Tab:parametrosGLC}
\end{table}

Finally, in our analysis we also will estimate the nuclear vector meson photoproduction cross - section considering that only the nucleons of the nuclear target that are not in the overlapping region  contribute for the interaction, i.e. that only spectator nucleons act as target. Following Ref. \cite{GayDucati:2018who}, in this case, we will assume
\begin{equation}
\sigma_{\gamma B \rightarrow V B}(W,b) = \int d{^2\rb_2}\, \theta(b_1 - R_A) \left\{ \int \int dz \, d^2\rr  [\psi^*_V(r,z)\psi(r,z)]_T\, \mathcal{N}_B(x, \rr, \rb_2) \right\}^2\,\,.
\label{Eq:sec_bdep}
\end{equation} 
It is important to emphazise that the theta function implies that $\sigma_{\gamma B \rightarrow V B}$ becomes dependent on $\rb$, since the relation  $\rb_1 = \rb + \rb_2$ (See Fig. \ref{colisao_geometria}) implies that \(b_1^2 = b^2 + b_2^2 + 2b b_2 \cos(\alpha)\).

\begin{figure}[t]
	\centering
\includegraphics[width=\linewidth]{sig_UPC_PC_Jpsi_50-90_IPsat2018_GLC_5.02TeV.eps} \\
\vspace{1.5cm}
    	\includegraphics[width=\linewidth]{real_Y_50-90_IPsat2018_GLC_5.02TeV.eps} \\
	\caption{Predictions for the rapidity distributions associated with the photoproduction of $J/\Psi$ (upper panels) and $\Upsilon$ (lower panels) mesons in peripheral $PbPb$ collisions at  \( \sqrt{s} = 5.02\,\mathrm{TeV} \)  and different centralities, derived  assuming distinct models for the effective nuclear photon flux and for the treatment of the photon - nucleus cross - section. The results were derived considering the IP-SAT model for the dipole - proton scattering amplitude and the GLC model for the overlap function. The experimental data is from ALICE Collaboration \cite{ALICE:2022zso,Massacrier:2024fgx,Bize:2024ros,ALICE:2024whv}.}
    \label{Fig:jpsi_difsecs}
\end{figure}


\section{Results}
\label{sec:results}
In this section we will present our predictions for the rapidity distributions associated with the photoproduction of $J/\Psi$ and $\Upsilon$ mesons in peripheral $PbPb$ collisions  at the LHC energies, considering dfferent centralities. We will consider distinct combinations of the models for the nuclear photon flux, photon - nucleus cross - section, overlap function and dipole - proton scattering amplitude. Our goal is to estimate the current theoretical uncertainty in the treatment of peripheral collisions. For a similar analysis in the case of ultraperipheral heavy - ion collisions  see, e.g. Ref. \cite{run2}. In addition, the average rapidity distribution in the kinematical range covered by the ALICE data will be estimated and a comparison with the current data will be performed. Following Refs. \cite{Klusek-Gawenda:2015hja, GayDucati:2018who}, the predictions for the distinct centralities will be estimated considering that the centrality $c$ and the impact parameter of the collision $b$ are related by  $c = {b^2}/({4R_A^2})$, where $R_A$ is the nuclear radius. { Two comments are in order. First, the relation between the centrality and the corresponding impact parameter of the collision, can also be estimated using a Monte Carlo based on the Glauber approach (See, e.g. Ref. \cite{Loizides:2017ack}), which imply different values for $b_{min}$ and $b_{max}$ in a given centrality range in comparison with the approach assumed here and in Refs. \cite{Klusek-Gawenda:2015hja, GayDucati:2018who}. 
In Appendix \ref{ssec:MC} we estimate the dependence of our predictions on this choice.
Second, we will focus our analysis on  large centralities (50\% - 70\% and 70\% - 90\%), but ALICE Collaboration has also released experimental data for smaller centralities ranges \cite{ALICE:2015mzu,ALICE:2022zso,ALICE:2024whv}. For completeness, a comparison between our predictions and these data is presented in Appendix \ref{ssec:lowcentra}. However, in our opinion, for centralities smaller than 50\%, the impact of the QGP formation on the photoproduced heavy vector mesons cannot be disregarded in order to derive a realistic prediction for the yields (See, e.g. Ref. \cite{Shi:2017qep}). As a consequence, the results presented in Appendix \ref{ssec:lowcentra} must be considered as upper bounds for the yields. A more detailed analysis, taking into account of the QGP effects, will be presented in a forthcoming publication. }

Initially, we will investigate the impact of the treatment of the photon - nucleus cross - section on the predictions of the rapidity distribution. In Fig. \ref{Fig:jpsi_difsecs}  we present our results for $J/\Psi$ (upper panels) and $\Upsilon$ (lower panels) production,  derived considering the IP-SAT model for the dipole - proton scattering amplitude and the GLC model for the overlap function. The current ALICE data are also presented for comparison \cite{ALICE:2022zso,Massacrier:2024fgx,Bize:2024ros}. One has that the predictions are strongly dependent on the modeling of the effective nuclear flux. In addition, the modification of $\sigma_{\gamma A}$, as performed in Eq. (\ref{Eq:sec_bdep}), implies a reduction of the predictions, that is larger for smaller centralities. Such a behavior is expected, since we are excluding the overlap region and considering only the spectators as target. Therefore, the predictions derived assuming the $b$ -dependent photon - nucleus cross - section, Eq. (\ref{Eq:sec_bdep}),  should be considered a lower bound for the photoproduction of vector mesons in peripheral collisions.
We also have verified that similar conclusions are derived considering  other models for 
${\cal N}_p$ and $(\Psi^{V*}\Psi)_{T}$, as well as for peripheral $PbPb$ collisions at \( \sqrt{s} = 2.76\,\mathrm{TeV} \).

In what follows, we will estimate the dependence of our predictions on the modeling of dipole - proton scattering amplitude and overlap function. For simplicity, the results will be derived  assuming the $b$
- dependent photon - nucleus cross - section, Eq. (\ref{Eq:sec_bdep}). 
In Fig. \ref{Fig:jpsiDSIGDY_rs5020} we present our predictions for the rapidity distributions associated with the photoproduction of $J/\Psi$ mesons in peripheral $PbPb$ collisions at  \( \sqrt{s} = 5.02\,\mathrm{TeV} \) and different centralities,  assuming distinct models for the effective nuclear photon flux,  dipole - proton scattering amplitude and  overlap function. One has that the BG model for the overlap function implies an increasing of the normalization, with the impact being larger for smaller centralities and for the IP-SAT and IPnon-SAT models. Moreover, for the centrality of $50 \% - 70 \%$, the $N^{(1)}$ ($N^{(2)}$) model for the effective nuclear photon flux provides an upper (lower) bound for the predictions. In contrast, for the centrality of $70 \% - 90 \%$, the lower bound is provided by the $N^{(0)}$ model. Regarding the dependence on 
${\cal N}_p$, the predictions associated with the bCGC model provide an upper bound at central rapidities, with the IP-SAT and IPnon-SAT results being similar. { The difference between the IP-SAT and IPnon-SAT increases at forward/backward rapidities, when smaller values of the Bjorken-$x$ variable are probed in comparison to midrapidities.}
 Although the comparison with the current data still does not allow us to discriminate between the different models, the results indicate that future experimental data for different centralities and rapidity ranges will provide important constraints on the modeling of peripheral collisions.

In Fig. \ref{Fig:YDSIGDY_rs5020} we present our results for the 
photoproduction of $\Upsilon$ mesons in peripheral $PbPb$ collisions.   We have that the dependencies on the modeling of overlap function, dipole - proton scattering amplitude and centralities are similar to those observed for the $J/\Psi$ case, differing in the magnitude of the normalization.

Finally,  in Tables  \ref{Tab:Jpsi} and \ref{Tab:Upsilon} we present our predictions for the average rapidity distributions, defined by
\begin{equation}
	\left\langle \frac{d\sigma^V}{d y} \right\rangle = \frac{1}{\Delta y} \int_{y_{min}}^{y_{max}} \frac{d\sigma(PbPb \rightarrow Pb  \otimes V \otimes Pb)}{d y} \, d y \,\,,
\end{equation}
where $\Delta y = y_{max} - y_{min}$. In order to compare our results with the ALICE data, we will assume $y_{min} = 2.5$ and $y_{max} = 4.0$. In particular, in  Table \ref{Tab:Jpsi}, we present our results for the $J/\Psi$ production  considering the distinct models for the effective nuclear photon flux, overlap function and dipole - proton scattering amplitude. The ALICE measurements are shown for comparison \cite{ALICE:2015mzu,ALICE:2022zso,ALICE:2024whv}. A stronger conclusion about the correctness (or not) of a given  model is still not  possible. As stated before, we believe that it will be possible with the analysis of the $\Upsilon$ production.  The corresponding results for the $\Upsilon$ production are presented in Table \ref{Tab:Upsilon}. { For completeness, in Appendix \ref{ssec:ultra} we present our UPC predictions.}

{ A final comment is in order. In our opinion, the simultaneous analysis of different heavy meson final states, such as the one presented here, will be fundamental to 
disentangle between the different models and assumptions used in the modelling of photoproduction processes in peripheral collisions.  Since all predictions were derived considering the same ingredients, future data will allow us to perform a complementary check of the  model predictions. In recent years, different authors have proposed the study of ratio between cross - sections for the vector meson photoproduction as a way to disentangle the description of the QCD dynamics \cite{Peredo:2023oym,Kovchegov:2023bvy}. Motivated by these results, we have estimated the ratios 
\(\left\langle\dd\sigma^{\Upsilon}/\dd y\right\rangle/\left\langle\dd\sigma^{J/\psi}/\dd y\right\rangle\) and
\(\left\langle\dd\sigma^{V}/\dd y\right\rangle_\text{peripheral}/\left\langle\dd\sigma^{V}/\dd y\right\rangle_\text{ultraperipheral}\). The corresponding results are presented in the Tables \ref{tab:ratiomesons}, 
\ref{tab:ratioPerUPC_jpsi} and \ref{tab:ratioPerUPC_ups}. Our results indicate that a future experimental analysis of these ratios could be very useful to discriminate between the different approaches for the description of the  effective nuclear photon flux,  dipole - proton scattering amplitude and  overlap functions. }


\begin{figure}[t]
	\centering
	\includegraphics[width=\linewidth]{1real_Jpsi_50-90_IPsat_nonsat_bCGC_BG_GLC_5.02TeV.eps}
	\caption{Predictions for the rapidity distributions associated with the photoproduction of $J/\Psi$ mesons in peripheral $PbPb$ collisions at  \( \sqrt{s} = 5.02\,\mathrm{TeV} \) and different centralities, derived assuming distinct models for the effective nuclear photon flux,  dipole - proton scattering amplitude and  overlap functions.  The results were obtained considering the that photon - nucleus cross - section is described by  Eq. (\ref{Eq:sec_bdep}). The experimental data is from ALICE Collaboration \cite{ALICE:2022zso,Massacrier:2024fgx,Bize:2024ros,ALICE:2024whv}.}
    \label{Fig:jpsiDSIGDY_rs5020}
\end{figure}

\begin{figure}[t]
	\centering
	\includegraphics[width=\linewidth]{real_Y_50-90_IPsat_nonsat_bCGC_BG_GLC_5.02TeV.eps}
\caption{Predictions for the rapidity distributions associated with the photoproduction of $\Upsilon$ mesons in peripheral $PbPb$ collisions at  \( \sqrt{s} = 5.02\,\mathrm{TeV} \) and different centralities,  derived assuming distinct models for the effective nuclear photon flux,  dipole - proton scattering amplitude and  overlap functions.  The results were obtained considering the that photon - nucleus cross - section is described by  Eq. (\ref{Eq:sec_bdep}).}
	\label{Fig:YDSIGDY_rs5020}
\end{figure}

\begin{table}[t]
	\centering
	\renewcommand{\arraystretch}{1.2}
	\begin{tabular}{||l|c|c|c|c||}
		\hline\hline
		\multicolumn{5}{c}{\(\langle{\dd\sigma^{J/\psi}}/{\dd y}\rangle~(\mu\textnormal{b})\)} \\
		\hline
				& \multicolumn{2}{c|}{\bf 2.76 TeV} 
		& \multicolumn{2}{c||}{\bf 5.02 TeV} \\
        \hline 
	{\bf Dipole - proton model}	& {\bf 50\%--70\%} & {\bf 70\%--90\%} & {\bf 50\%--70\%} & {\bf 70\%--90\%} \\
		\hline\hline
		
		\multicolumn{5}{c}{ \text{\bf Effective nuclear photon flux: \( N^{(0)}(\omega,b) \)}} \\
        \hline 
bCGC         & {99.21}\,({92.97}) & {66.50}\,({62.94}) & {219.84}\,({200.13}) & {156.68}\,({143.32}) \\
IP-SAT   & {95.59}\,({82.48}) & {63.74}\,({55.78}) & {216.83}\,({179.25}) & {153.77}\,({127.96}) \\
IPnon-SAT    & {104.09}\,({90.19}) & {69.26}\,({60.92}) & {238.86}\,({197.57}) & {169.21}\,({140.94}) \\
\hline
\hline
		\multicolumn{5}{c}{ \text{\bf Effective nuclear photon flux: \( N^{(1)}(\omega,b) \)}} \\
        \hline
bCGC         & {183.67}\,({169.57}) & {105.95}\,({98.85}) & {344.51}\,({311.86}) & {220.34}\,({200.49}) \\
IP-SAT   & {178.67}\,({150.92}) & {102.41}\,({87.81}) & {342.10}\,({280.57}) & {217.63}\,({179.72}) \\
IPnon-SAT    & {195.49}\,({165.55}) & {111.77}\,({96.17}) & {377.76}\,({309.84}) & {240.04}\,({198.32}) \\
		\hline
\hline		
		\multicolumn{5}{c}{ \text{\bf Effective nuclear photon flux: \( N^{(2)}(\omega,b) \)}} \\
        \hline
bCGC         & {87.33}\,({81.58}) & {82.40}\,({77.27}) & {186.05}\,({169.23}) & {179.23}\,({163.41}) \\
IP-SAT   & {84.30}\,({72.43}) & {79.42}\,({68.58}) & {183.70}\,({151.68}) & {176.60}\,({146.26}) \\
IPnon-SAT    & {91.92}\,({79.26}) & {86.54}\,({75.04}) & {202.44}\,({167.24}) & {194.62}\,({161.28}) \\
\hline
		\hline
				\multicolumn{5}{c}{ \text{\bf Effective nuclear photon flux: \( N^{(3)}(\omega,b) \)}} \\
                \hline
bCGC         & {99.97}\,({93.39}) & {86.11}\,({80.75}) & {212.98}\,({193.72}) & {187.27}\,({170.75}) \\
IP-SAT   & {96.53}\,({82.92}) & {82.99}\,({71.67}) & {210.29}\,({173.63}) & {184.53}\,({152.83}) \\
IPnon-SAT    & {105.22}\,({90.72}) & {90.44}\,({78.42}) & {231.74}\,({191.44}) & {203.36}\,({168.53}) \\
\hline
\hline
        {\bf ALICE data $(2.5 \le y \le 4.0)$} & {\( 58 \pm 16^{+8}_{-10} \pm 8 \)} & {\( 59 \pm 11^{+7}_{-10} \pm 8 \)} & {\( 216 \pm 10 \pm 12 \)} & {\( 167 \pm 6 \pm 12 \)} \\
		\hline\hline
	\end{tabular}
	\caption{Predictions for the average rapidity distribution associated with the photoproduction of $J/\Psi$ mesons in  peripheral $PbPb$ collisions at \( \sqrt{s} = 2.76\,\mathrm{TeV} \) and \( \sqrt{s} = 5.02\,\mathrm{TeV} \) and different centralities,  derived considering distinct models for the effective nuclear photon flux,  dipole - proton scattering amplitude and overlap function. The results in parentheses are for the GLC model.
 For completeness, the experimental results measured by the ALICE Collaboration are also presented \cite{ALICE:2015mzu,ALICE:2022zso}.}
    \label{Tab:Jpsi}
\end{table}

\begin{table}[t]
	\centering
	\renewcommand{\arraystretch}{1.2}
	\begin{tabular}{||l|c|c||}
		\hline\hline
		\multicolumn{3}{c}{\(\langle{\dd\sigma^{\Upsilon}}/{\dd y}\rangle~(\textnormal{nb})\)} \\
		\hline 
		& \multicolumn{2}{c||}{\bf 5.02 TeV} \\
        \hline
{\bf Dipole proton model}		& {\bf 50\%--70\%} & {\bf 70\%--90\%} \\
		\hline\hline
		\multicolumn{3}{c}{ \text{\bf Effective nuclear photon flux: \( N^{(0)}(\omega,b) \)}} \\
        \hline
bCGC         & {125.08} \,({107.04}) & {71.37}\,({61.87}) \\
IP-SAT  & {194.13}\,({159.07}) & {109.50}\,({90.75}) \\
IPnon-SAT    & {198.01}\,({163.42}) & {111.21}\,({92.83}) \\
		\hline 
        \hline
		\multicolumn{3}{c}{ \text{\bf Effective nuclear photon flux: \( N^{(1)}(\omega,b) \)}} \\
        \hline
bCGC         & {344.12}\,({289.13}) & {157.89}\,({134.04}) \\
IP-SAT   & {549.67}\,({442.66}) & {248.43}\,({202.00}) \\
IPnon-SAT    & {564.45}\,({458.12}) & {254.18}\,({208.26}) \\
		\hline
        \hline
		\multicolumn{3}{c}{ \text{\bf Effective nuclear photon flux: \( N^{(2)}(\omega,b) \)}} \\
        \hline
bCGC         & {121.96}\,({103.82}) & {109.90}\,({93.89}) \\
IP-SAT   & {190.80}\,({155.57}) & {171.51}\,({140.25}) \\
IPnon-SAT    & {194.99}\,({160.16}) & {175.09}\,({144.25}) \\
		\hline
        \hline
		\multicolumn{3}{c}{ \text{\bf Effective nuclear photon flux: \( N^{(3)}(\omega,b) \)}} \\
        \hline
bCGC         & {139.71}({118.93}) & {114.88}\,({98.14}) \\
IP-SAT   & {218.59}\,({178.22}) & {179.29}\,({146.62}) \\
IPnon-SAT    & {223.39}\,({183.48}) & {183.04}\,({150.80}) \\
		\hline\hline
	\end{tabular}
	\caption{Predictions for the average rapidity distribution associated with the photoproduction of $\Upsilon$ mesons in  peripheral $PbPb$ collisions at  \( \sqrt{s} = 5.02\,\mathrm{TeV} \) and different centralities,  derived considering distinct models for the effective nuclear photon flux,  dipole - proton scattering amplitude and overlap function. The results in parentheses are for the GLC model.}
	\label{Tab:Upsilon}
\end{table}

\begin{table}[t]
	\centering
	\renewcommand{\arraystretch}{1.2}
	\begin{tabular}{||l|c|c||}
		\hline\hline
		\multicolumn{3}{c}{%
			\rule[-3ex]{0pt}{7.5ex}%
			\(\dfrac{\langle{\dd\sigma^{\Upsilon}}/{\dd y}\rangle~(\textnormal{nb})}{\langle{\dd\sigma^{J/\psi}}/{\dd y}\rangle~(\mu\textnormal{b})}\)} \\
		\hline
		{\bf Dipole - proton model} &  {\bf 50\%--70\%} & {\bf 70\%--90\%} \\
		\hline\hline
		
		\multicolumn{3}{c}{ \text{\bf Effective nuclear photon flux: \( N^{(0)}(\omega,b) \)}} \\
		\hline
		bCGC    &  0.00057\,(0.00053) & 0.00046\,(0.00043) \\
		IP-SAT  &    0.00090\,(0.00089) & 0.00071\,(0.00071) \\
		IPnon-SAT &   0.00083\,(0.00083) & 0.00066\,(0.00066) \\
		\hline\hline
		
		\multicolumn{3}{c}{ \text{\bf Effective nuclear photon flux: \( N^{(1)}(\omega,b) \)}} \\
		\hline
		bCGC  &       0.00100\,(0.00093) & 0.00072\,(0.00067) \\
		IP-SAT &     0.00161\,(0.00158) & 0.00114\,(0.00112) \\
		IPnon-SAT &  0.00149\,(0.00148) & 0.00106\,(0.00105) \\
		\hline\hline
		
		\multicolumn{3}{c}{ \text{\bf Effective nuclear photon flux: \( N^{(2)}(\omega,b) \)}} \\
		\hline
		bCGC     &   0.00066\,(0.00061) & 0.00061\,(0.00057) \\
		IP-SAT    &  0.00104\,(0.00103) & 0.00097\,(0.00096) \\
		IPnon-SAT &  0.00096\,(0.00096) & 0.00090\,(0.00089) \\
		\hline\hline
		
		\multicolumn{3}{c}{ \text{\bf Effective nuclear photon flux: \( N^{(3)}(\omega,b) \)}} \\
		\hline
		bCGC      &  0.00066\,(0.00061) & 0.00061\,(0.00057) \\
		IP-SAT    &  0.00104\,(0.00103) & 0.00097\,(0.00096) \\
		IPnon-SAT  &  0.00097\,(0.00096) & 0.00090\,(0.00090) \\
		\hline\hline
	\end{tabular}
	\caption{Ratio between the average rapidity distributions for \(\Upsilon\) and \(J/\psi\) meson photoproduction in peripheral  \(PbPb\) collisions at \(5.02\,\mathrm{TeV}\), considering different effective nuclear photon fluxes, dipole–proton models and wave functions. Values in parentheses correspond to the GLC wave function.}
	\label{tab:ratiomesons}
\end{table}

\begin{table}[t]
	\centering
	\renewcommand{\arraystretch}{1.2}
	\begin{tabular}{||l|c|c|c|c||}
		\hline\hline
		\multicolumn{5}{c}{\(\left\langle\dd\sigma^{J/\psi}/\dd y\right\rangle_\text{peripheral}/\left\langle\dd\sigma^{J/\psi}/\dd y\right\rangle_\text{ultraperipheral}\)} \\
		\hline
		& \multicolumn{2}{c|}{\bf 2.76 TeV} & \multicolumn{2}{c||}{\bf 5.02 TeV} \\
		\hline 
		{\bf Dipole - proton model} & {\bf 50\%--70\%} & {\bf 70\%--90\%} & {\bf 50\%--70\%} & {\bf 70\%--90\%} \\
		\hline\hline
		
		\multicolumn{5}{c}{\text{\bf Effective nuclear photon flux: \( N^{(0)}(\omega,b) \)}} \\
		\hline
		bCGC       & 0.0921\,(0.0831) & 0.0617\,(0.0563) & 0.1001\,(0.0915) & 0.0713\,(0.0655) \\
		IP-SAT     & 0.0960\,(0.0828) & 0.0640\,(0.056)  & 0.1094\,(0.0958) & 0.0776\,(0.0684) \\
		IPnon-SAT  & 0.0987\,(0.0839) & 0.0657\,(0.0567) & 0.1132\,(0.0980) & 0.0802\,(0.0699) \\
		\hline\hline
		
		\multicolumn{5}{c}{\text{\bf Effective nuclear photon flux: \( N^{(1)}(\omega,b) \)}} \\
		\hline
		bCGC       & 0.1705\,(0.1516) & 0.0983\,(0.0883) & 0.1568\,(0.1425) & 0.1003\,(0.0916) \\
		IP-SAT     & 0.1795\,(0.1515) & 0.1029\,(0.0882) & 0.1726\,(0.1500) & 0.1098\,(0.096) \\
		IPnon-SAT  & 0.1854\,(0.1540) & 0.1060\,(0.0926) & 0.1790\,(0.1537) & 0.1138\,(0.0984) \\
		\hline\hline
		
		\multicolumn{5}{c}{\text{\bf Effective nuclear photon flux: \( N^{(2)}(\omega,b) \)}} \\
		\hline
		bCGC       & 0.0811\,(0.0729) & 0.0762\,(0.0691) & 0.0847\,(0.0773) & 0.0815\,(0.0747) \\
		IP-SAT     & 0.0847\,(0.0727) & 0.0798\,(0.0689) & 0.0927\,(0.0811) & 0.0891\,(0.0782) \\
		IPnon-SAT  & 0.0872\,(0.0737) & 0.0821\,(0.0698) & 0.0959\,(0.0830) & 0.0922\,(0.0799) \\
		\hline\hline
		
		\multicolumn{5}{c}{\text{\bf Effective nuclear photon flux: \( N^{(3)}(\omega,b) \)}} \\
		\hline
		bCGC       & 0.0928\,(0.0835) & 0.0799\,(0.0722) & 0.0969\,(0.0885) & 0.0853\,(0.0781) \\
		IP-SAT     & 0.0970\,(0.0832) & 0.0833\,(0.0719) & 0.1061\,(0.0928) & 0.0931\,(0.0818) \\
		IPnon-SAT  & 0.0998\,(0.0844) & 0.0858\,(0.073)  & 0.1098\,(0.0950) & 0.0964\,(0.0836) \\
		\hline\hline
	\end{tabular}
	\caption{Ratio between the average rapidity distributions for \(J/\psi\) meson photoproduction in peripheral and ultraperipheral \(PbPb\) collisions at \(\sqrt{s} = 2.76\,\mathrm{TeV}\) and \(5.02\,\mathrm{TeV}\), considering different effective nuclear photon fluxes, dipole–proton models and wave functions. Values in parentheses correspond to the GLC wave function.}
	\label{tab:ratioPerUPC_jpsi}
	\end{table}

\begin{table}[t]
	\centering
	\renewcommand{\arraystretch}{1.2}
	\begin{tabular}{||l|c|c||}
		\hline\hline
		\multicolumn{3}{c}{\(\left\langle\dd\sigma^{\Upsilon}/\dd y\right\rangle_\text{peripheral}/\left\langle\dd\sigma^{\Upsilon}/\dd y\right\rangle_\text{ultraperipheral}\)} \\
		\hline 
		{\bf Dipole - proton model}	& \textbf{50--70\%} & \textbf{70--90\%} \\
		\hline\hline
		\multicolumn{3}{c}{\textbf{Effective nuclear photon flux: \( N^{(0)}(\omega,b) \)}} \\
		\hline
		bCGC & 0.2055 (0.1766) & 0.1172 (0.1021) \\
		IP-SAT & 0.2317 (0.1986) & 0.1307 (0.1133) \\
		IPnon-SAT & 0.2442 (0.2095) & 0.1372 (0.1190) \\
		\hline\hline
		\multicolumn{3}{c}{\textbf{Effective nuclear photon flux: \( N^{(1)}(\omega,b) \)}} \\
		\hline
		bCGC & 0.5653 (0.4769) & 0.2594 (0.2211) \\
		IP-SAT & 0.6561 (0.5528) & 0.2966 (0.2522) \\
		IPnon-SAT & 0.6962 (0.5874) & 0.3135 (0.2670) \\
		\hline\hline
		\multicolumn{3}{c}{\textbf{Effective nuclear photon flux: \( N^{(2)}(\omega,b) \)}} \\
		\hline
		bCGC & 0.2003 (0.1712) & 0.1805 (0.1549) \\
		IP-SAT & 0.2278 (0.1943) & 0.2047 (0.1751) \\
		IPnon-SAT & 0.2405 (0.2053) & 0.2160 (0.1849) \\
		\hline\hline
		\multicolumn{3}{c}{\textbf{Effective nuclear photon flux: \( N^{(3)}(\omega,b) \)}} \\
		\hline
		bCGC & 0.2295 (0.1962) & 0.1887 (0.1619) \\
		IP-SAT & 0.2609 (0.2225) & 0.2140 (0.1831) \\
		IPnon-SAT & 0.2755 (0.2352) & 0.2258 (0.1933) \\
		\hline\hline
	\end{tabular}
	\caption{Ratio between the average rapidity distributions for \(\Upsilon\) meson photoproduction in peripheral and ultraperipheral \(PbPb\) collisions at \(\sqrt{s} = 5.02\,\mathrm{TeV}\), considering different effective nuclear photon fluxes, dipole--proton models and wave functions. Values in parentheses correspond to the GLC wave function.}
		\label{tab:ratioPerUPC_ups}
\end{table}

\section{Summary}
\label{sec:summary}

In the last years, experimental results obtained by the STAR and ALICE Collaborations for the dilepton and vector meson production have demonstrated that the contribution of the  $\gamma \gamma \rightarrow l^+l^-$ and $\gamma A \rightarrow V A$ processes are not negligible, especially when the final state has a very low transverse momentum. However, the description of these photon - induced interactions in peripheral heavy ion collisions is still an open question. In this paper, we have performed a comprehensive study of the photoproduction of heavy vector mesons in peripheral $PbPb$ collisions at the LHC energies. In particular, we have considered distinct approaches for the main ingredients of the calculation in order to estimate the current theoretical uncertainty. Our goal was to compare the results of these distinct approaches with the current $J/\Psi$ data and provide predictions for the $\Upsilon$ productions. The results indicated that a future global analysis of the data for both mesons will be very useful to improve our understanding of photon - induced interactions in peripheral collisions.

\begin{acknowledgments}
V.P.G. acknowledges useful discussions with   Mariola Klusek - Gawenda, Sony Martins and Antoni Szczurek over the last years, and is grateful to the Mainz Institute of Theoretical Physics (MITP) of the Cluster of Excellence PRISMA+ (Project ID 390831469), for its hospitality and  support.  V.P.G. was partially supported by CNPq,  FAPERGS and INCT-FNA (Process No. 464898/2014-5). P.E.A.C and B.D.M. were partially supported by CAPES and FAPESC. A.V.G.is grateful to Universidade do Estado de Santa Catarina for its hospitality and financial support. The authors acknowledge the National Laboratory for Scientific Computing (LNCC/MCTI, Brazil), through the ambassador program (UFGD), subproject FCNAE for providing HPC 
resources of the SDumont supercomputer.

\end{acknowledgments}

\newpage
\appendix

\section{Predictions for small centralities}
\label{ssec:lowcentra}
As discussed in Section \ref{sec:results}, 
for centralities smaller than 50\%, the impact of the QGP formation on the photoproduced heavy vector mesons is expected to be non negligible and, in principle, must be considered in order to derive a realistic prediction for the yields (See, e.g. Ref. \cite{Shi:2017qep}). However, as the ALICE Collaboration has also released experimental data for smaller centralities ranges \cite{ALICE:2015mzu,ALICE:2022zso,ALICE:2024whv}, in Figs. \ref{Fig:Jpsi_4050} and \ref{Fig:Jpsi_1050} we present our predictions for the rapidity distributions associated with the photoproduction of $J/\Psi$ mesons in peripheral $PbPb$ collisions at  \( \sqrt{s} = 5.02\,\mathrm{TeV} \) and  centralities 40\% - 50\%
and 10\% - 50\%, respectively. The corresponding predictions for the average rapidity distribution associated with the photoproduction of $J/\Psi$ mesons in  peripheral $PbPb$ collisions at \( \sqrt{s} = 2.76\,\mathrm{TeV} \) and \( \sqrt{s} = 5.02\,\mathrm{TeV} \) and centralities 10\% - 30\% and 30\% - 50\% are presented in Table \ref{tab:jpsicentral}. The ALICE data are also shown for comparison.
In our opinion, such results  must be considered as upper bounds for the yields. As already emphasized in Section \ref{sec:results}, a more detailed analysis, taking into account of the QGP effects, will be presented in a forthcoming publication.

\begin{figure}[t]
	\centering
	\includegraphics[width=\linewidth]{real_Jpsi_40-50_IPsat_nonsat_bCGC_BG_GLC_5.02TeV.eps}
\caption{Predictions for the rapidity distributions associated with the photoproduction of $J/\Psi$ mesons in peripheral $PbPb$ collisions at  \( \sqrt{s} = 5.02\,\mathrm{TeV} \) and  centralities 40\% - 50\%,  derived assuming distinct models for the effective nuclear photon flux,  dipole - proton scattering amplitude and  overlap functions.  The results were obtained considering the that photon - nucleus cross - section is described by  Eq. (\ref{Eq:sec_bdep}). The experimental data is from ALICE Collaboration \cite{ALICE:2024whv}}
	\label{Fig:Jpsi_4050}
\end{figure}

\begin{figure}[t]
	\centering
	\includegraphics[width=\linewidth]{1real_Jpsi_10-50_IPsat_nonsat_bCGC_BG_GLC_5.02TeV.eps}
\caption{Predictions for the rapidity distributions associated with the photoproduction of $J/\Psi$ mesons in peripheral $PbPb$ collisions at  \( \sqrt{s} = 5.02\,\mathrm{TeV} \) and centralities 10\% - 30\% and 30\% - 50\%,  derived assuming distinct models for the effective nuclear photon flux,  dipole - proton scattering amplitude and  overlap functions.  The results were obtained considering the that photon - nucleus cross - section is described by  Eq. (\ref{Eq:sec_bdep}). The experimental data is from ALICE Collaboration \cite{ALICE:2022zso}.}
	\label{Fig:Jpsi_1050}
\end{figure}

\begin{table}[t]
	\centering
	\renewcommand{\arraystretch}{1.2}
	\begin{tabular}{||l|c|c|c|c||}
		\hline\hline
		\multicolumn{5}{c}{\(\langle{\dd\sigma^{J/\psi}}/{\dd y}\rangle~(\mu\textnormal{b})\)} \\
		\hline
		& \multicolumn{2}{c|}{\bf 2.76 TeV} 
		& \multicolumn{2}{c||}{\bf 5.02 TeV} \\
		\hline 
		{\bf Dipole - proton model}	& {\bf 10\%--30\%} & {\bf 30\%--50\%} & {\bf 10\%--30\%} & {\bf 30\%--50\%} \\
		\hline\hline
		
		\multicolumn{5}{c}{ \text{\bf Effective nuclear photon flux: \( N^{(0)}(\omega,b) \)}} \\
		\hline 
		bCGC         & {299.68}\,({275.96}) & {157.98}\,({146.55}) & {519.45}\,({470.69}) & {319.6}\,({289.72}) \\
		IP-SAT   & {292.20}\,({245.41}) & {153.11}\,({130.22}) & {517.23}\,({423.79}) & {316.86}\,({260.34}) \\
		IPnon-SAT    & {321.55}\,({270.64}) & {167.3}\,({142.73}) & {574.38}\,({470.72}) & {349.81}\,({287.50}) \\
		\hline
		\hline
		\multicolumn{5}{c}{ \text{\bf Effective nuclear photon flux: \( N^{(1)}(\omega,b) \)}} \\
		\hline
		bCGC         & {816.36}\,({743.87}) & {410.84}\,({375.14}) & {1146.01}\,({1033.62}) & {651.88}\,({587.33}) \\
		IP-SAT   & {802.51}\,({663.99}) & {402.7}\,({334.87}) & {1147.36}\,({934.1}) & {651.16}\,({530.45}) \\
		IPnon-SAT    & {886.07}\,({733.96}) & {442.36}\,({368.43}) & {1276.27}\,({1038.97}) & {720.81}\,({587.12}) \\
		\hline
		\hline		
		\multicolumn{5}{c}{ \text{\bf Effective nuclear photon flux: \( N^{(2)}(\omega,b) \)}} \\
		\hline
		bCGC         & {49.11}\,({45.76}) & {77.41}\,({72.11}) & {100.65}\,({91.62}) & {161.97}\,({147.08}) \\
		IP-SAT   & {47.51}\,({40.61}) & {74.85}\,({64.04}) & {99.58}\,({82.17}) & {160.17}\,({131.98}) \\
		IPnon-SAT    & {52.02}\,({44.62}) & {81.68}\,({70.13}) & {110.21}\,({91}) & {176.58}\,({145.64}) \\
		\hline
		\hline
		\multicolumn{5}{c}{ \text{\bf Effective nuclear photon flux: \( N^{(3)}(\omega,b) \)}} \\
		\hline
		bCGC         & {89.03}\,({82.96}) & {103.85}\,({96.74}) & {182.31}\,({166.03}) & {217.13}\,({197.23}) \\
		IP-SAT   & {86.14}\,({73.62}) & {100.42}\,({85.91}) & {180.39}\,({148.87}) & {214.74}\,({176.96}) \\
		IPnon-SAT    & {94.34}\,({80.90}) & {109.58}\,({94.09}) & {199.71}\,({164.91}) & {236.83}\,({195.27}) \\
		\hline
		\hline
		{\bf ALICE data} $(2.5 \le y \le 4.0)$ & {\(< 290 \)} & {\(  73 \pm 44^{+26}_{-27} \pm 10 \)} & {\( 145 \pm 62 \pm 85 \)} & {\( 179 \pm 24 \pm 22 \)} \\
		\hline\hline
	\end{tabular}
\caption{Predictions for the average rapidity distribution associated with the photoproduction of $J/\Psi$ mesons in  peripheral $PbPb$ collisions at \( \sqrt{s} = 2.76\,\mathrm{TeV} \) and \( \sqrt{s} = 5.02\,\mathrm{TeV} \) and different centralities,  derived considering distinct models for the effective nuclear photon flux,  dipole - proton scattering amplitude and overlap function. The results in parentheses are for the GLC model. For completeness, the experimental results measured by the ALICE Collaboration are also presented \cite{ALICE:2015mzu,ALICE:2022zso}.}
\label{tab:jpsicentral}
\end{table}

\section{Dependence on the relation between the centrality and the impact parameter of the collision}
\label{ssec:MC}
Following Refs. \cite{Klusek-Gawenda:2015hja, GayDucati:2018who}, in our analysis we have considered that the relation between the centrality $c$ and the impact parameter  of the collision $b$ is given by 
$c = {b^2}/({4R_A^2})$, where $R_A$ is the nuclear radius. However, such a relation can also 
be established using a  Monte Carlo based on the Glauber approach (See, e.g. Ref. \cite{Loizides:2017ack}). Such distinct approaches imply distinct values for the minimum and maximum impact parameters associated with a given centrality range. Consequently, these two approaches imply different predictions for peripheral heavy ion collisions and have different impact for distinct  effective nuclear photon fluxes. The results presented in Fig. \ref{Fig:MC}, for two models of the effective nuclear photon flux, demonstrate this dependence. We have that the predictions associated with the Glauber Monte Carlo calculation \cite{Loizides:2017ack}, denoted by $b_{MCG}$, are larger than those derived using the geometrical approach ($b_{geometric}$), with the difference being larger for smaller centralities. Such a difference motivates the improving of the current geometrical approaches  considered for the treatment of peripheral collisions.

\begin{figure}[t]
	\centering
	\includegraphics[width=\linewidth]{MCG_real_Jpsi_50-90_IPsat2018_GLC_5.02TeV.eps}
\caption{Predictions for the rapidity distributions associated with the photoproduction of $J/\Psi$ mesons in peripheral $PbPb$ collisions at  \( \sqrt{s} = 5.02\,\mathrm{TeV} \) and different centralities,  derived assuming distinct models for  the relation between the centrality and the impact parameter.  The results were obtained considering the that photon - nucleus cross - section is described by  Eq. (\ref{Eq:sec_bdep}). The experimental data is from ALICE Collaboration \cite{ALICE:2022zso,Massacrier:2024fgx,Bize:2024ros,ALICE:2024whv}.}
	\label{Fig:MC}
\end{figure}

\section{Predictions for ultraperipheral $PbPb$ collisions}
\label{ssec:ultra}
For completeness of our analysis, in Tables \ref{tab:uphicjpsi} and \ref{tab:uphicups}, we present our predictions for the average rapidity distribution associated with the photoproduction of $J/\psi$ and $\Upsilon$ mesons in  ultraperipheral $PbPb$ collisions,  derived considering distinct models for the dipole - proton scattering amplitude and overlap function. As expected from the analysis performed  in Ref. \cite{run2}, the predictions are dependent on the models assumed for the dipole - target cross - section and overlap functions.

\begin{table}[t]
	\centering
	\renewcommand{\arraystretch}{1.2}
	\begin{tabular}{||l|c|c||}
		\hline\hline
		\multicolumn{3}{c}{\(\langle{\dd\sigma^{J/\psi}}/{\dd y}\rangle~(\mu\textnormal{b})\)} \\
		\hline
		{\bf Dipole proton model}		& {\bf 2.76 TeV} & {\bf 5.02 TeV} \\
		\hline
		bCGC         & {1077.26}\,({1118.70}) & {2197.08}\,({2188.09}) \\
		IP-SAT  & {995.64}\,({996.20}) & {1982.20}\,({1870.80}) \\
		IPnon-SAT    & {1054.44}\,({1075.14}) & {2110.23}\,({2016.00}) \\
		\hline 
		\hline
	\end{tabular}
	\caption{Predictions for the average rapidity distribution associated with the photoproduction of $J/\psi$ mesons in  ultraperipheral $PbPb$ collisions at  \( \sqrt{s} = 2.76\,\mathrm{TeV} \) and \( \sqrt{s} = 5.02\,\mathrm{TeV} \),  derived considering distinct models for the dipole - proton scattering amplitude and overlap function. The results in parentheses are for the GLC model.}
	\label{tab:uphicjpsi}
\end{table}

\begin{table}[t]
	\centering
	\renewcommand{\arraystretch}{1.2}
	\begin{tabular}{||l|c||}
		\hline\hline
		\multicolumn{2}{c}{\(\langle{\dd\sigma^{\Upsilon}}/{\dd y}\rangle~(\textnormal{nb})\)} \\
		\hline
		{\bf Dipole proton model}		&  {\bf 5.02 TeV} \\
		\hline
		bCGC         & {608.79}\,({606.26})  \\
		IP-SAT  &  {810.73}\,({779.97})  \\
		IPnon-SAT    & {837.73}\,({800.82})  \\
		\hline 
		\hline
	\end{tabular}
	\caption{Predictions for the average rapidity distribution associated with the photoproduction of $\Upsilon$ mesons in  ultraperipheral $PbPb$ collisions at  \( \sqrt{s} = 5.02\,\mathrm{TeV} \),  derived considering distinct models for the dipole - proton scattering amplitude and overlap function. The results in parentheses are for the GLC model.}
\label{tab:uphicups}
\end{table}

\bibliographystyle{unsrt}

\end{document}